\documentclass[conference]{IEEEtran}
\IEEEoverridecommandlockouts
\usepackage{cite}
\usepackage{amsmath,amssymb,amsfonts}
\usepackage{algorithmic}
\usepackage{graphicx}
\usepackage{textcomp}
\usepackage{xcolor}
\usepackage{url}
\usepackage{amsmath}
\usepackage{amssymb}

\newcommand{\etal}{\textit{et al. }}

\newcommand{\orig}{x}
\newcommand{\pred}{x_\mathrm{p}}

\newcommand{\latentC}{\tilde{x}_\mathrm{p}}
\newcommand{\res}{r}

\newcommand{\bit}{\,\mathrm{bit}}
\newcommand{\wrong}{w}
\newcommand{\prob}{p}
\newcommand{\sigmap}{\sigma_\mathrm{p}}

\def\BibTeX{{\rm B\kern-.05em{\sc i\kern-.025em b}\kern-.08em
    T\kern-.1667em\lower.7ex\hbox{E}\kern-.125emX}}
\begin{document}

\title{On\,Benefits\,and\,Challenges\,of\,Conditional\,Interframe Video\,Coding\,in\,Light\,of\,Information\,Theory
}

\author{\IEEEauthorblockN{Fabian Brand, J\"urgen Seiler, and Andr\'e Kaup}
\IEEEauthorblockA{\textit{Multimedia Communications and Signal Processing} \\
\textit{Friedrich-Alexander-Universit\"at Erlangen-N\"urnberg}\\
Erlangen, Germany \\
\texttt{\{fabian.brand,juergen.seiler,andre.kaup\}@fau.de}}
}

\maketitle

\makeatletter
\def\ps@IEEEtitlepagestyle{%
	\def\@oddfoot{\mycopyrightnotice}%
	\def\@oddhead{\hbox{}\@IEEEheaderstyle\leftmark\hfil\thepage}\relax
	\def\@evenhead{\@IEEEheaderstyle\thepage\hfil\leftmark\hbox{}}\relax
	\def\@evenfoot{}%
}
\def\mycopyrightnotice{%
	\begin{minipage}{\textwidth}
		\scriptsize
		\copyright 2022 IEEE.  Personal use of this material is permitted. Permission from
		IEEE must be obtained for all other uses, in any current or future
		media, including reprinting/republishing this material for advertising
		or promotional purposes, creating new collective works, for resale or
		redistribution to servers or lists, or reuse of any copyrighted
		component of this work in other works.
	\end{minipage}
}
\makeatother

\begin{abstract}
The rise of variational autoencoders for image and video compression has opened the door to many elaborate coding techniques. One example here is the possibility of conditional interframe coding. Here, instead of transmitting the residual between the original frame and the predicted frame (often obtained by motion compensation), the current frame is transmitted under the condition of knowing the prediction signal. In practice, conditional coding can be straightforwardly implemented using a conditional autoencoder, which has also shown good results in recent works. In this paper, we provide an information theoretical analysis of conditional coding for inter frames and show in which cases gains compared to traditional residual coding can be expected. We also show the effect of information bottlenecks which can occur in practical video coders in the prediction signal path due to the network structure, as a consequence of the data-processing theorem or due to quantization. We demonstrate that conditional coding has theoretical benefits over residual coding but that there are cases in which the benefits are quickly canceled by small information bottlenecks of the prediction signal.
\end{abstract}

\begin{IEEEkeywords}
video compression, inter coding, conditional coding, information theory, conditional autoencoder
\end{IEEEkeywords}

\section{Introduction}
The invention of autoencoders for image compression~\cite{BalleLS2017_Endendoptimized} has shaped the research in this area decisively for the last few years. Similar to existing image and video compression standards like JPEG~\cite{Wallace1992_JPEGstillpicture}, JPEG2000~\cite{ITUTI2004_JPEG2000Image}, HEVC~\cite{SullivanOH2012_OverviewHighEfficiency}, VVC~\cite{BrossCL2020_VersatileVideoCoding}, AV1~\cite{HanLM2021_TechnicalOverviewAV1}, or many more, this method relies on a transformation of the image into a sparse or low-dimensional domain. Different to the traditional methods, transforms in learning-based compression are non-linear and data-driven. Typically, these transforms are implemented as convolutional neural networks which are trained as a constrained autoencoder. 

Learning-based approaches have also been proposed for video compression. The general structure of the approaches has largely been taken over from known hybrid video coding solutions, such as VVC or AV1. At first, motion is estimated, then the motion field is transmitted, typically in a lossy way. Afterwards motion compensation is performed. This yields a prediction frame which is used to reduce the temporal redundancy between frames. The task at hand is now to transmit the remaining information to obtain the reconstructed frame. In most published learning-based approaches~\cite{LuOX2018_DVCEndend,AgustssonMJ2020_ScaleSpaceFlow,HuLX2021_FVCNewFramework}, the strategy is taken over from traditional video compression, i.e., the difference between the original frame and the prediction frame, the so-called residual signal, is computed and transmitted. At the decoder, the reconstructed residual is added to the prediction signal to obtain the reconstructed frame. This strategy has been proven efficient in both traditional video coding and also in deep-learning-based video coding. However, the latter case opens the door to a different and more general possibility: Conditional coding. In conditional coding, we do not compress the residual, but rather the frame itself under the condition of knowing a prediction. Using neural networks, conditional coders can be implemented using a conditional autoencoder. This structure extends the autoencoder by adding an additional signal, the condition, to both encoder and decoder. The network therefore learns by itself how to exploit the redundancy between the signals.

Conditional coding exploits a relationship from information theory: The entropy of the difference is greater or equal to the conditional entropy of the signal $\orig$ given its prediction $\pred$:
\begin{equation}
H(\orig - \pred) \ge H(\orig|\pred)
\end{equation}
This inequality suggests that in theory conditional coding is always at least as good as conventional residual coding. 

In this theoretical work, we examine the possibility of conditional coding for video compression in light of information theory. The equation above suggests conditional coding is always superior to residual coding. In the first part of this paper, we go one step further and precisely derive the theoretically possible gain of conditional coding over residual coding. We furthermore take possible information bottlenecks into account, which may occur in practical implementations during the processing of the prediction signal, and we give bounds how such bottlenecks influence the coding performance. In the second part of the paper we validate our theoretical results in simulations. We gain more insights into when conditional coding is beneficial and where its limits are in practical scenarios.

\section{Related Work}
All modern video coders make use of inter prediction to reduce the temporal redundancy between frames. Block-based coders such as HEVC~\cite{SullivanOH2012_OverviewHighEfficiency}, VVC~\cite{BrossCL2020_VersatileVideoCoding}, VP9~\cite{MukherjeeBG2013_latestopensource} or AV1~\cite{HanLM2021_TechnicalOverviewAV1} estimate motion on a block-level, yielding a prediction signal per block. If no suitable block is available in the reference frame, the coders have the possibility to locally switch back to intra prediction using previously decoded content. After prediction, a residual is computed by subtracting the prediction signal from the original. The resulting residual is then transformed using a handcrafted frequency transform such as the discrete cosine transform, a discrete sine transform or a combination thereof. Furthermore, AV1 and VVC allow dynamic switching between different transforms.

In 2019, the deep video compression framework (DVC)~\cite{LuOX2018_DVCEndend} was published. This work introduced the first end-to-end trained video coder consisting of motion estimation, motion transmission, and residual compression. Additionally, DVC contained a network to enhance the prediction frame. For motion estimation, a SpyNet~\cite{RanjanB2017_OpticalFlowEstimation} architecture is used. Both motion and residual compression use a standard autoencoder, similar to~\cite{BalleLS2017_Endendoptimized}. In subsequent work, the feature-space video coder (FVC)~\cite{HuLX2021_FVCNewFramework} shifts motion estimation and compensation into the feature space. Here, the framework also uses residual coding to transmit the residual features.

\begin{figure}
	\centering
	\begin{tabular}{cc}
		\includegraphics[height=0.6\columnwidth]{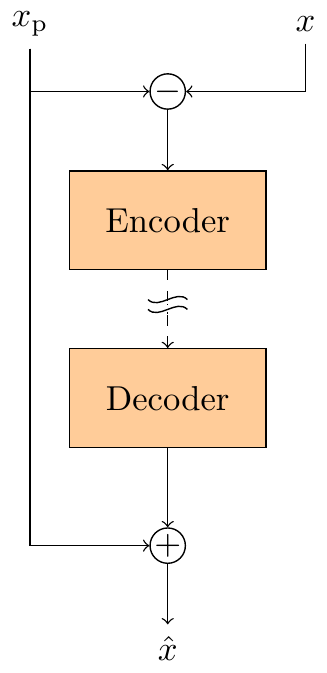} &
		\includegraphics[height=0.6\columnwidth]{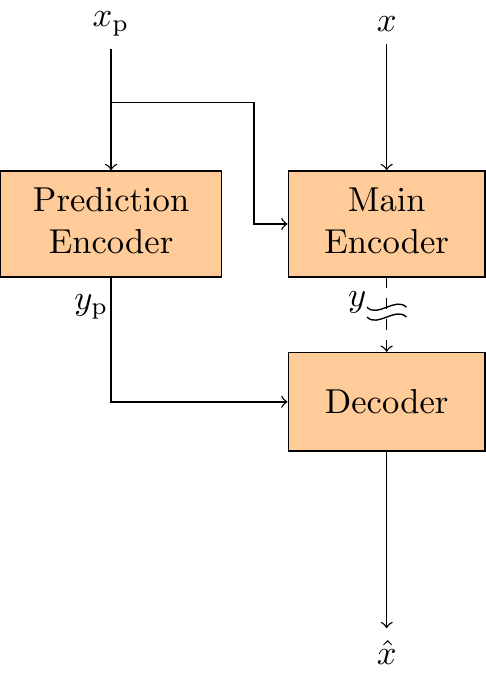}\\
		Difference Coder & Conditional Coder from\\& \cite{LadunePH2020_ModeNetModeSelection} and \cite{LadunePH2020_OpticalFlowMode}\\
	\end{tabular}
	\caption{Left: Conventional residual coder. Right: Schematic network structure of conditional autoencoder as proposed as \textit{CodecNet} in \cite{LadunePH2020_ModeNetModeSelection} and \cite{LadunePH2020_OpticalFlowMode}. Dashed lines denote transmission over a channel.}\label{Fig:SchematicsLadune}\vspace{-0.5cm}
\end{figure}

DVC and FVC still follow the basic paradigm of residual coding, taken from traditional inter coding approaches. In \cite{LadunePH2020_ModeNetModeSelection} and subsequently in~\cite{LadunePH2020_OpticalFlowMode,LadunePH2021_ConditionalCodingVariable}, Ladune \etal proposed CodecNet, a conditional coding approach, together with auxiliary networks named ModeNet or MOFNet, which enable a skip mode. Skip modes are commonly found in traditional codecs and describe copying of the prediction signal directly into the reconstructed frame without residual transmission. As shown in Fig. \ref{Fig:SchematicsLadune}, the conditional coding approach used in CodecNet consists of two encoders. One conditional branch encoder which generates a latent representation $y_\mathrm{p}$ of the prediction signal $\pred$ and a main encoder, which has both the prediction signal and the original signal $x$ as input.
The main encoder compresses the original signal under the condition of knowing the prediction signal. The resulting latent representation $y$ is coded and transmitted over the channel. The decoder uses the compressed latent representation of the original and the latent representation of the prediction signal to reconstruct the original frame. Note that the latent representation of the prediction signal is not transmitted since it can be constructed at the decoder. The decoder therefore reconstructs the frame under the condition of knowing \emph{a latent representation} of the prediction signal.

In previous work, we proposed the extended generalized difference coder (xGDC)~\cite{BrandSK2022_PFrameCoding}, which is another instance of a conditional coder. Here, we used trainable networks to replace difference and sum in traditional residual coding. This generalized residual signal is then compressed jointly with the traditional linear residual. Additionally the encoder signals which part of the image is to be reconstructed with the conditional coder and which part with the residual coder.

\section{Residual Coding versus Difference Coding}
\subsection{Ideal Conditional Coding}
\label{Sec:Theory}
In order to get a better understanding of residual coding in video compression, we want to consider the different scenarios in the light of information theory. The Shannon entropy serves as a lower bound of the bitrate needed to compress a signal. In the following, we assume ideal lossless coders which reach entropy. Let $H(x)$ denote the entropy of the distribution from which $x$ is drawn. So, if $x$ symbolizes a natural image, $H(x)$ is the entropy of natural images. In residual coding, we compress the residual $r=\!\orig\!-\!\pred$ and in conditional coding, we compress the original frame $\orig$ under the condition of knowing $\pred$. We therefore have to compare the entropy of residual frames $H(r) = H(\orig - \pred)$ and the conditional entropy $H(\orig|\pred)$. We start by using Bayes law, to derive
\begin{equation}
H(\orig,\pred|\res) + H(\res) = H(\orig,\pred,\res) = H(\orig,\pred) + \underbrace{H(\res|\orig,\pred)}_{=0}.
\end{equation}
We can easily see that the conditional entropy $H(\res|\orig,\pred)$ must be zero, since the residual is completely determined from knowing $\orig$ and $\pred$.
We can therefore continue by summarizing and rearranging:
\begin{equation}
\begin{split}
H(\res) &= H(\orig,\pred) - H(\orig,\pred|\res) \\
&=H(\orig|\pred) + H(\pred) - \underbrace{H(\orig|\pred,\res)}_{=0} - H(\pred|\res) \\
&=H(\orig|\pred) + H(\pred) - H(\pred|\res) = H(\orig|\pred) + I(\pred,\res).
\end{split}
\end{equation}
In the second line, we can see that $H(\orig|\pred,\res)$ must be zero because the original frame $\orig$ can be reconstructed from the prediction signal $\pred$ and the residual $\res$. $H(\pred) - H(\pred|\res)$ is the mutual information $I(\pred,\res)$ and hence we can write:
\begin{equation}
H(\orig-\pred) = H(\orig|\pred) + I(\pred;\res)
\label{Eq:Main}
\end{equation}
Since the mutual information is non-negative, this implies
\begin{equation}
H(\orig-\pred) \geq H(\orig|\pred),
\end{equation}
with equality if and only if $\res$ and $\pred$ have no mutual information.

From this inequality we obtain general insights about the efficiency of conditional coding compared to residual coding. We not only see that conditional coding is (in theory) at least as good as residual coding but we can also quantify how large the difference is. The larger the mutual information between the residual and the prediction frame, the larger is the gain of conditional coding. 

\subsection{Information Bottlenecks}
When interpreting the results, we need to take into account possible bottlenecks between the prediction signal and the output. In this context, a bottleneck describes a loss of information content in the prediction signal before the reconstruction of the frame. It is easy to see that no bottleneck occurs in residual coding. Here that prediction signal is directly used to reconstruct the original frame. In conditional coding, however, the prediction signal has to be processed, so there are multiple possibilities where bottlenecks may occur in the processing chain. 

One example where such a bottleneck appears is the conditional autoencoder proposed in \cite{LadunePH2020_ModeNetModeSelection} and \cite{LadunePH2020_OpticalFlowMode}. The schematic structure of this coder is given in Fig.~\ref{Fig:SchematicsLadune}. There, an additional encoder is used to obtain a latent representation $y_\mathrm{p}$ of the prediction signal $\pred$, which is then used together with the transmitted latent representation $y$ to reconstruct the frame. Hence, the decoder only sees a latent representation of the prediction signal, which can not contain the full information of the prediction signal. 

Other ways for bottlenecks to appear in the processing chain come from the data processing theorem~\cite{ViterbiO1979_PrinciplesDigitalCommunication}. This theorem states that every processing on a signal will either decrease the entropy of that signal or keep it the same. Therefore every processing step has the potential to reduce the information content in a non-reversible way. Since (finite-size) convolutions are in general non-reversible (by finite-size convolutions), each convolutional layer processing the prediction signal potentially reduces the information content. Additionally, it is easy to see that ReLU non-linearities also are prone to reducing entropy of the prediction.

Another bottleneck arises in practical scenarios when we consider quantized operations. In order to obtain speed and bit-exact reproducibility, which are requirements of real-world coders, all calculations have to be performed in fixed-point arithmetic. Ball\'e \etal showed in~\cite{BalleJM2019_IntegerNetworksData} that it is possible to integerize networks, however this comes at a cost of reduced precision. Preserving the information of the prediction signal poses another challenge to integer networks.

We can model any bottleneck with a general function $\latentC = f\left(\pred\right)$, where $\latentC$ is a degraded version of $\pred$. It is clear that
\begin{equation}
H(\pred) \ge H(\latentC)
\label{Eq:Entropy}
\end{equation}
and
\begin{equation}
H(\orig|\pred) \le H(\orig|\latentC)
\label{Eq:CondEntropy}
\end{equation}
hold. Furthermore, it can be shown that 
\begin{equation}
H(\orig|\pred) = H(\orig|\latentC) - I(\orig;\pred|\latentC)
\label{Eq:CondEntropy2}
\end{equation}
Plugging this result into (\ref{Eq:Main}), we obtain
\begin{equation}
H(\orig-\pred) = H(\orig|\latentC) - I(\orig;\pred|\latentC) + I(\pred;\res).
\label{Eq:Main2}
\end{equation}
Since $I(\orig;\pred|\latentC)$ is non-negative, $H(\orig\!-\!\pred)\geq H(\orig|\latentC)$ does not necessarily hold, depending on how much information is lost during the bottleneck $f$.

\section{Analysis}

In the following section we analyze the behavior of the entropy in a simplified inter coding scenario. Since there are no accurate probability models for entire pictures available, computing the precise entropy of an image is not possible. 

\subsection{Setup}
We consider the following scenario: We only consider one pixel of an image near an object boundary. Let the value of this pixel be $x$ and let $x$ be uniformly distributed between 0 and 255. It is easy to see that without further information $H(x) = 8\bit$. Furthermore, let $\wrong$ be another uncorrelated pixel value from the reference frame which lies at the other side of the object boundary. We model our prediction signal $\pred$ for $x$ as follows: with a probability of $(1-\prob)$, $x$ is predicted correctly, i.e. $\pred=x$ and with probability $\prob$, the deviation in motion vector causes a prediction from $\wrong$, i.e. $\pred=\wrong$.  The overall conditional probability mass function of the predictor is therefore:
\begin{equation}
	f_{X_\mathrm{p}|X}(\pred|x) = \prob\cdot\delta[\pred-\wrong] + (1-\prob)\cdot\delta[\pred-x],
\end{equation}
with the discrete Dirac impulse $\delta(x)$.


To model the degradation by a bottleneck occurring during conditional coding we uniformly quantize our prediction signal to 7\,bit or 6\,bit. Note that in an actual coder, the bottleneck might not occur in the pixel domain but in some latent representation. However, quantization is able to reduce the information content of the effective prediction signal, and has therefore the same effect as a bottleneck. We denote the degraded prediction signal as $\latentC$.

During our experiments, we compare the entropies \mbox{$H(x\!-\!\pred)$}, representing traditional residual coding, $H(x|\pred)$ representing idealized conditional coding not taking into account bottlenecks, and $H(x|\latentC)$, which is a more realistic approximation of conditional coding. Our goal is to show that idealized conditional coding never performs worse than residual coding and in most cases clearly outperforms the traditional method. Second, we want to show in which cases potential bottlenecks have a large effect, introducing challenges for conditional coding.

\subsection{Results}
\begin{figure}
	\includegraphics[width=\linewidth]{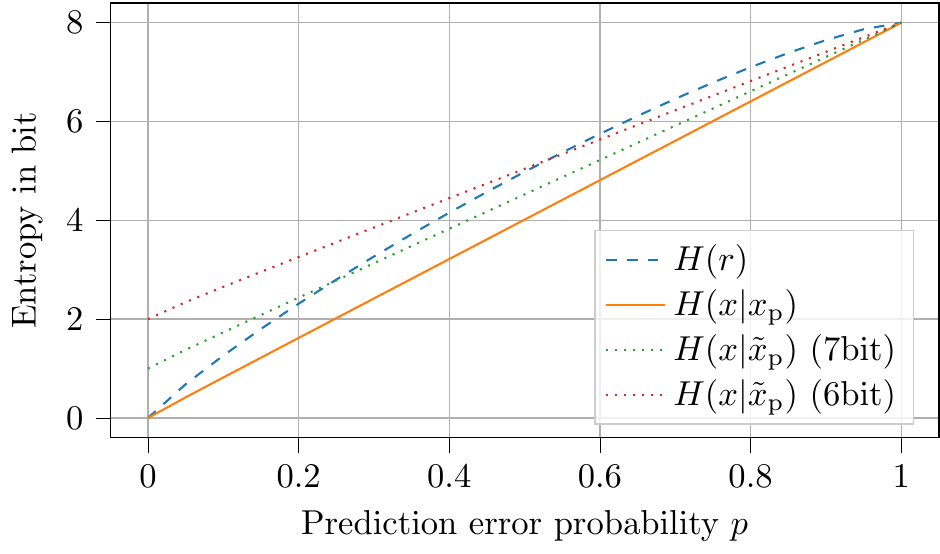}
	\caption{Entropy of difference coding and conditional coding. We show conditional coding with no bottleneck and with 7 and 6\,bit bottleneck.\label{Fig:Prob}}
\end{figure}

We first examine the influence of the prediction error probability $\prob$. In Fig.~\ref{Fig:Prob} the blue and orange curve show $H(\res)$ and $H(x|\pred)$, respectively, when we vary the probability $\prob$ of a wrong reference pixel. We can see that $H(x|\pred) = H(r)$ only for $\prob = 0$ and $\prob = 1$. For all other cases, $H(\res)$ is strictly greater than $H(x|\pred)$. Interestingly, the curve of $H(x|\pred)$, almost looks like a linear function in $\prob$. When we look at the closed form expression
\begin{equation}
	H(x|\pred) = \frac{N\prob+1}{N+1}\log_2\left(N\prob+1\right)-\frac{N\prob}{N+1}\log_2p,
\end{equation}
with $N = 2^8-1=255$ in this case, one can prove that it converges toward a linear function for $N\rightarrow\infty$. On the other hand, we see that $H(\res)$ is a concave function with the same beginning and end point as $H(x|\pred)$. 

When we add a bottleneck in the conditional coder, we need to transmit additional information. In general, we see that we need more additional information when we have a tighter bottleneck.With increasing error probability, $H(x|\latentC)$ gets closer to $H(\res)$ until eventually the conditional coder performs better again. 

From this we can already gain some intuitions, which we will confirm later with more experiments. Bottlenecks in conditional coding increase the necessary rate. This was expected, since the information content of the prediction signal decreases. When the quality of the prediction decreases (in this case due to motion inaccuracy) and the rate increases, this effect is getting smaller and the loss due to the bottleneck decreases. We see that the curves with bottleneck approach the ideal conditional coder. Then the transmitted signal itself contains more information and does not rely on the prediction signal that much, therefore inaccuracies in the prediction signal are compensated better. We also observe that the conditional coding gain depends on the prediction error probability $\prob$.

\begin{figure}
	\includegraphics[width=\columnwidth]{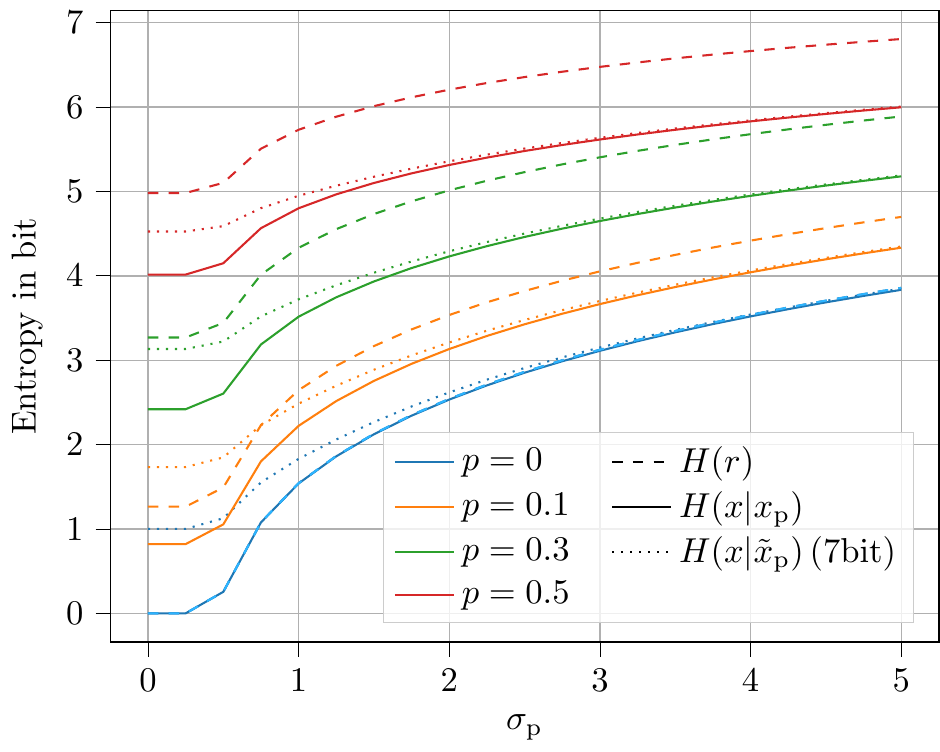}
	\caption{Entropies of conditional and residual coding when we vary $\sigmap$. We show the results for multiple values of $\prob$.\label{Fig:Sigma}}
\end{figure}

A confusion with a non-correlated pixel due to motion inaccuracy is only one of the possible errors which can occur during inter prediction. Even if the pixel was predicted from the correct reference pixel, present uncorrelated noise in both frames can lead to prediction inaccuracies. Similarly, in highly correlated areas of an image, displacement errors can be modeled as additive noise. For further reading on this topic see~\cite{Girod1987_EfficiencyMotionCompensating}. We model the error with additive Gaussian noise with zero mean and standard deviation $\sigmap$. In Fig.~\ref{Fig:Sigma}, we examine how conditional coding behaves when we vary $\sigmap$. 

Here, we observe a key difference from the first experiment. As expected, both $H(r)$ and $H(x|\pred)$ increase with increasing prediction error $\sigmap$. However, the gap between the curves is constant over $\sigmap$. Recall that in the previous experiment, the gain depended on $\prob$ and thereby on the prediction quality. This can be explained when we look at $\prob=0$. Here, we observe no conditional coding gain. Recall that the conditional coding gain is equal to the mutual information between $\res$ and $\pred$, $I(\res; \pred)$. We see that since $\res$ in this case is uncorrelated noise, this mutual information must be zero. Therefore, this results fits our theoretical results from the previous section. 

We also show the effects of bottlenecks in this scenario. In this case we observe the case for a 7\,bit bottleneck. We see that especially for small $\sigmap$, the presence of a bottleneck increases the entropy greatly. However as $\sigmap$ increases, the curves more and more align. This result again shows that bottlenecks are most relevant in areas with very good predictors.

\begin{figure}
	\begin{tabular}{c@{\hspace{2mm}}cc@{\hspace{2mm}}c}
		\includegraphics[width=0.2\columnwidth, clip, trim=100 100 350 0]{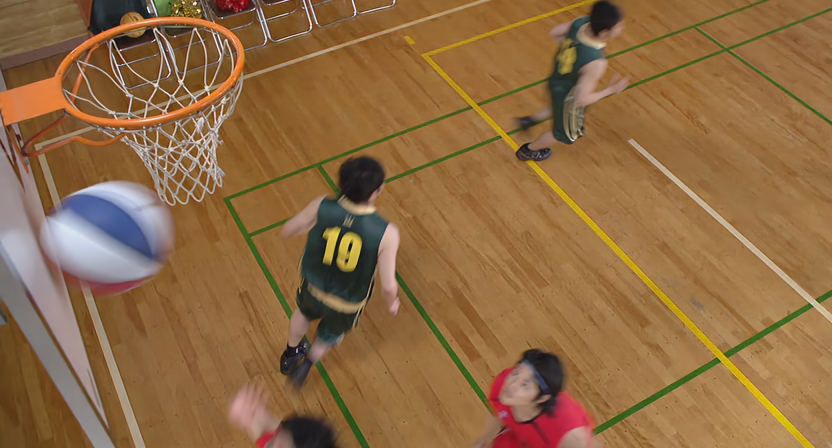} &
		\includegraphics[width=0.2\columnwidth, clip, trim=100 100 350 0]{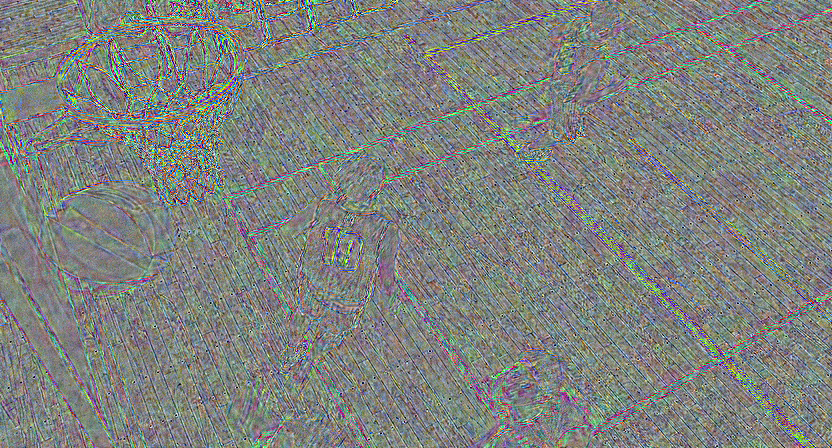} & 
		\includegraphics[width=0.2\columnwidth, clip, trim=100 100 350 0]{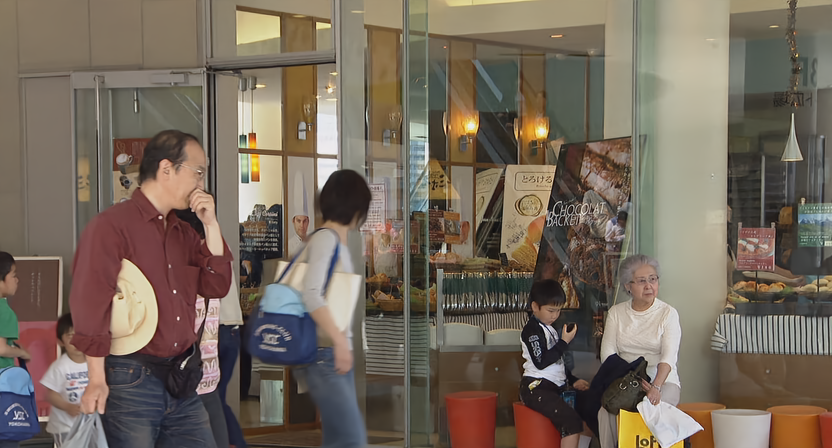} &
		\includegraphics[width=0.2\columnwidth, clip, trim=100 100 350 0]{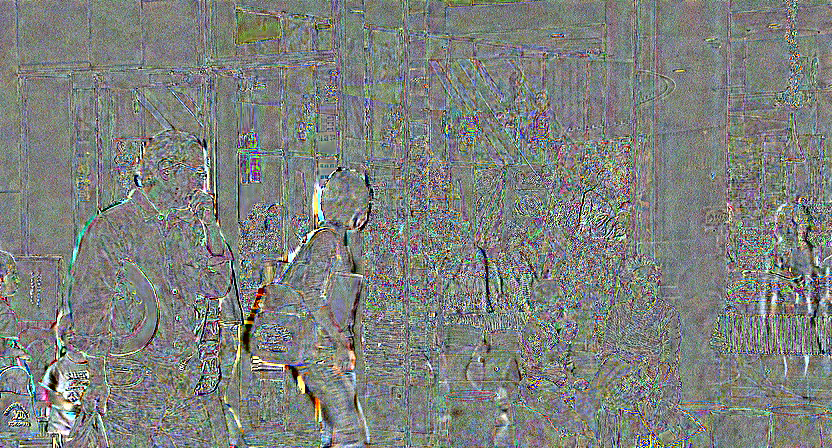}\\
		Prediction $\pred$ & Residual $\res$ & Prediction $\pred$ & Residual $\res$
	\end{tabular}
	\caption{Example to demonstrate the mutual information of the prediction signal $\pred$ and the residual signal $\res$ (Here displayed with increased amplitude and shifted such that gray represents zero). The prediction in this example was computed using the optical flow from SpyNet~\cite{RanjanB2017_OpticalFlowEstimation}.\label{Fig:Example}}
	\vspace{-0.5cm}
\end{figure}

Finally, we want to demonstrate that $I(r;\pred)$ is greater than zero for real images. Since computing the precise mutual information for images is not possible, we show examples of a prediction signal and the corresponding residual, in Fig.~\ref{Fig:Example}. It is obvious that we can see structures of the prediction signal in the residual, especially around object boundaries, where small motion vector errors have a larger effect. Since thereby obviously contains information,  $I(r;\pred)$ is strictly greater zero, which can be exploited by conditional coding.

\subsection{Benefits and Challenges}
In the previous subsection, we clearly showed the potential of conditional coding to perform better than residual coding. A downside of conditional coding is the need to process the data further, with can yield inadvertent information bottlenecks, which we also consider in our experiments. 

From our computations and simulations, we arrive at the following benefits:
\begin{itemize}
	\item Under ideal conditions conditional coding never performs worse than residual coding.
	\item Even with moderate bottlenecks, conditional coding performs better when the prediction signal has small errors.
\end{itemize}

On the other hand, we also identified challenges  when designing conditional interframe coders:

\begin{itemize}
	\item Information bottlenecks between the prediction and the reconstruction signal must be avoided.
	\item For low prediction quality, the gain of the ideal conditional coder is possible also with bottlenecks, for high quality, residual coding may be better. Content adaptive coding is therefore desirable.
\end{itemize}

When we look at conditional inter coders that were published before, we see confirmation for our theoretical deliberations. Ladune~\etal~\cite{LadunePH2020_OpticalFlowMode} showed that conditional coding performs better than residual coding. Additionally, their method employs an additional network called ModeNet, which bypasses the conditional coder and copies the prediction signal directly into the reconstructed signal. This yields a behavior similar to skip modes in traditional coders. In our previous work~\cite{BrandSK2022_PFrameCoding}, we proposed xGDC, which contains a switch between conditional coding and residual coding, thereby also bypassing the bottleneck, also beating a comparable residual coder. In both approaches, the conditional coder is bypassed for areas with very good prediction signals. This validates our theoretical assessment which calls for a strategy to avoid bottlenecks.

\section{Conclusion}
In this paper we examined the theoretical properties of conditional inter coding compared to residual coding. Conditional coding solves the general task of transmitting a frame when a prediction signal is known. Residual compression, is one special case of conditional coding, therefore general conditional coding is always as least as good as residual coding. In this work, we computed the gain of conditional coding compared to residual coding to be the mutual information of the residual and the prediction signal, $I(\pred, \res)$. We showed this relationship in multiple experiments. 

We additionally showed that conditional coding can create information bottlenecks between the prediction and the reconstructed signal, which can impair the quality of conditional coding, particularly for very good prediction quality and low rate. We showed theoretical results how large the losses due to bottlenecks can be and showed in our simulations when this effect is particularly large. We finally outlined the benefits and challenges of conditional coding compared to residual coding and showed that our theoretical deliberations explain results found in published conditional coding approaches. 

In this paper we showed the large potential of conditional coding. While residual coding, a low complexity approximation of true conditional coding was dominant for a long time, using end-to-end trained neural networks conditional is now possible. This paper provides theoretical foundations for this coding approach and hopefully encourages further research into the efficient methods how conditional coding concepts can be applied in practice.
\bibliographystyle{IEEEtran}

\end{document}